\documentclass[preprint,12pt]{aastex}
\usepackage{natbib}
\usepackage{amsmath}
\usepackage{epsfig}
\begin{document}

\title{HI Observations of the Supermassive Binary Black Hole System in 0402+379}

\author{ {C. Rodriguez\altaffilmark{1}}, {G. B. Taylor\altaffilmark{1,2}},
{R. T. Zavala\altaffilmark{3}},{Y. M. Pihlstr\"om,\altaffilmark{1,2}}, {A. B. Peck\altaffilmark{4,5}} }

\altaffiltext{1}{Department of Physics and Astronomy, University of New Mexico, Albuquerque, NM 87131}
\altaffiltext{2}{Greg Taylor and Ylva Pihlstr\"om are also Adjunct Astronomers at the National Radio Astronomy Observatory}
\altaffiltext{3}{United States Naval Observatory, Flagstaff Station 10391 W. Naval Observatory Rd. Flagstaff, AZ 86001}
\altaffiltext{4}{Joint ALMA Office, Avda El Golf 40, piso 18, Santiago, Chile}
\altaffiltext{5}{National Radio Astronomy Observatory, 520 Edgemont Rd, Charlottesville, VA 22903}

\begin{abstract}

We have recently discovered a supermassive binary black hole system with a projected separation between the two black holes of 7.3 parsecs in the radio galaxy 0402+379 \citep{Rod06}. This is the most compact supermassive binary black hole pair yet imaged by more than two orders of magnitude. We present Global VLBI observations at 1.3464 GHz of this radio galaxy, taken to improve the quality of the HI data.
Two absorption lines are found toward the southern jet of the source, one redshifted by 370 $\pm$ 10 km s$^{-1}$ and the other blueshifted by 700 $\pm$ 10 km s$^{-1}$ with respect to the systemic velocity of the source, which, along with the results obtained for the opacity distribution over the source, suggests the presence of two mass clumps rotating around the central region of the source.
We propose a model consisting of a geometrically thick disk, of which we only see a couple of clumps, that reproduces the velocities measured from the HI absorption profiles. These clumps rotate in circular Keplerian orbits around an axis that crosses one of the supermassive black holes of the binary system in 0402+379. We find an upper limit for the inclination angle of the twin jets of the source to the line of sight of $\theta=66^{\circ}$, which, according to the proposed model, implies a lower limit on the central mass of $\sim7 \times 10^8$ $M_{\odot}$ and a lower limit for the scale height of the thick disk of $\sim12$ pc .

\end{abstract}

\keywords{galaxies: active -- galaxies: individual (0402+379) -- 
radio continuum: galaxies -- radio lines: galaxies}

\section{Introduction}
\label{intro}

Given that most galaxies harbor supermassive black holes at their centers (Richstone et al. 1998; Gebhardt et al. 2000), and that galaxy mergers are common, binary black holes should likewise be common. An understanding of the evolution and formation of these systems is important for an understanding of the evolution and formation of galaxies in general (Silk $\&$ Rees 1998, Merritt 2006). Theoretical descriptions of supermassive binary black hole systems and their accretion disks have been investigated by Hayasaki et al. 2007, 2008 and by MacFadyen $\&$ Milosavljevi{\'c} 2008.

Our ability to resolve the supermassive black holes in any given binary system depends on the separation between them, on their distance from Earth, and on the resolving power of the telescope used. It is believed that the longest timescale in the evolution of a supermassive binary black hole system leading up to coalescence is the stage in which the system is closely bound ($\sim$ 0.1 - 10 pc), meaning that in most of these systems the black hole pair can only be resolved by VLBI observations (in the case where both black holes are radio loud) which provides resolutions of milliarcseconds and finer. This could explain why very few such systems have been found (see review by \cite{Komo03a} detailing observational evidence for supermassive black hole binaries). 

Some source properties like X-shaped radio galaxies and double-double radio galaxies, helical radio-jets, double-horned emission line profiles, and semi-periodic variations in lightcurves have been taken as indirect evidence for compact binary black holes though other explanations are possible. Some wider systems have, however, been found more directly. For example, the ultra luminous galaxy NGC 6240, discovered by the Chandra X-ray observatory, was found to have a pair of active supermassive black holes at its center \citep{Komo03b}, separated by a distance of 1.4 kpc. Another system that has been known for some time is the double AGN (7 kpc separation) constituting the radio source 3C 75, which was discovered by the VLA to have two pairs of radio jets \citep{Owen85}.

The radio galaxy 0402+379 was recently found to contain two central, compact, flat spectrum, variable components (designated C1 and C2), with a projected separation of 7.3 pc, a feature which had not been observed in any other compact source, making this system the most compact supermassive binary black hole pair yet imaged by more than two orders of magnitude, with an estimated system mass of a few $10^8$ $M_{\odot}$ (Maness et al. 2004; Rodriguez et al. 2006). 

\cite{Maness04} performed spectral line observations of 0402+379 at 1.348 GHz. HI absorption at a redshift of 560 km s$^{-1}$ from the systemic velocity of the source (16,489 $\pm$ 300 km s$^{-1}$; Xu et al. 1994) was observed and attributed to a high velocity gas system, possibly due to a merger. Following this discovery, a spectrum of the HI region was taken in 2004 with the Westerbork telescope \citep{Morganti}, which  showed two components with velocities separated by 1000 km s$^{-1}$. 
In this article we present Global VLBI observations taken to improve the quality of the HI data on 0402+379, with the purpose of  better understanding the origin of the deviation from the systemic velocity seen in the HI gas in this source.

At the redshift of 0402+379 of 0.055, H$_{0}$=75 km s$^{-1}$ Mpc$^{-1}$ and q$_{0}$ = 0.5, a scale of 1 mas = 1.06 pc is obtained.

\section{Observations}

\subsection{Global VLBI Observations} 

\label{Observations}

Global VLBI observations\footnote{Global VLBI = Very Long Baseline Array + Effelsberg + Jodrell Bank + Westerbork + Green Bank Telescope + Onsala.} were made on 2007 March 17 at 1.3464 GHz. A single intermediate frequency with a bandwidth of 16 MHz was observed with 256 channels in both right and left circular polarizations, resulting in a frequency resolution of 62.5 kHz, corresponding to a velocity resolution of 15 km s$^{-1}$. Four level quantization was employed. The net integration time on 0402+379 was 497 minutes.

Standard flagging, amplitude calibration, fringe-fitting, and bandpass calibration (3C 111 was used for both gain and bandpass calibration) were followed in the Astronomical Image Processing System (AIPS; van Moorsel et al. 1996). AIPS reduction scripts described in \citet{Ulvestad01} were used for a large part of the reduction. Spectral line Doppler corrections were also applied in AIPS. All manual editing, frequency averaging procedures, imaging, deconvolution, and self-calibration were done using Difmap \citep{Shep95}. A clean cube was also produced by first performing a continuum subtraction using the task UVLSF in AIPS, then using Difmap in order to obtain a cleaned map for each spectral frequency, and finally combining all the maps into a cube with the task MCUBE in AIPS.

\section{Results}

\subsection{Radio Continuum}
\label{RadioCont}

\begin{figure}
\centering
\scalebox{1}{\includegraphics{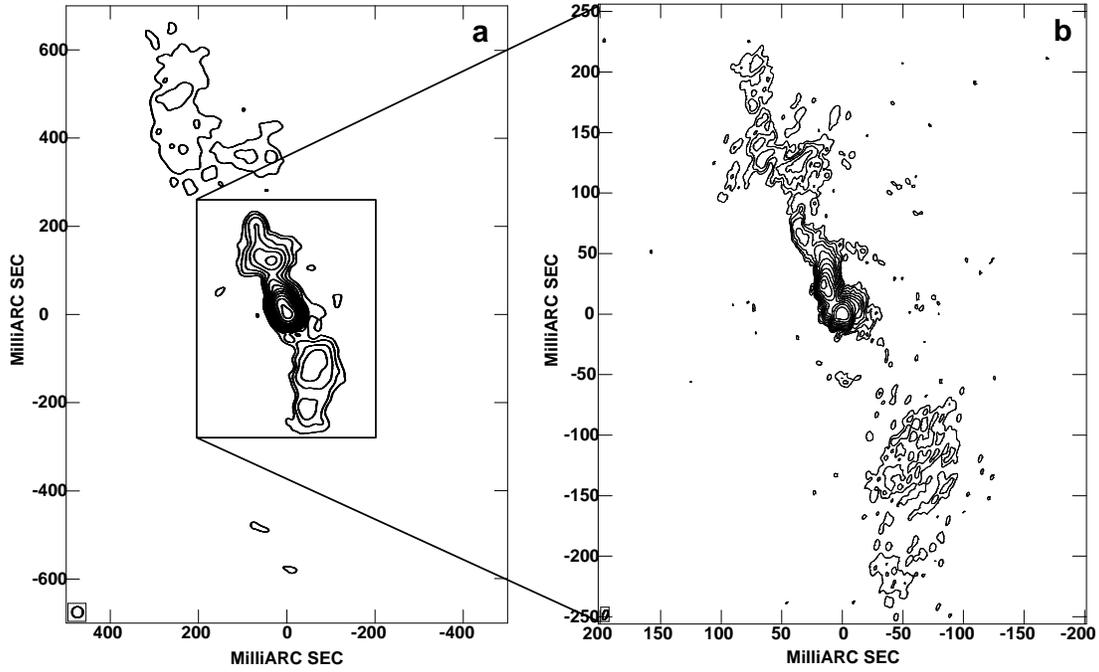}}
\vspace{2cm}
\caption{Naturally weighted 2007 Global VLBI image of 0402+379 at 1.3 GHz. Contours are drawn beginning at 3$\sigma$ and increase
by factors of 2 thereafter. 
In (b) the synthesized beam is 8.15 $\times$ 3.72 mas (shown in the bottom left corner). The peak flux density is 0.13 Jy/beam and rms noise is 0.02 mJy. In (a) the image was tapered and restored with a circular 25 mas synthesized beam (shown in the bottom left corner). The peak flux density is 0.46 Jy/beam and rms noise is 0.05 mJy.}
\label{1.3continuum}
\end{figure}

Figure \ref{1.3continuum} shows a naturally weighted 1.3 GHz image of 0402+379 from the 2007 Global VLBI observations. The source  consists of two diametrically opposed jets and a central region containing two active nuclei not resolved at this frequency (see Rodriguez et al. 2006 for a detailed multi-frequency study of this radio galaxy). The northern jet is pointing in the northeast direction whereas the southern jet is pointing in the southwest direction. The orientation of the source at this frequency is consistent with that seen by the VLA at both 1.5 and 5 GHz \citep{Maness04} and by the VLBA at 0.3 GHz \citep{Rod06}. The image shown in panel (a) was tapered and restored with a circular 25 mas synthesized beam, in order to show better the extended structure in this radio galaxy. The source spans $\sim1000$ mas ($\sim1000$ pc). The image shown in panel (b) was tapered and restored with a 8.15 $\times$ 3.72 mas synthesized beam. We see structure on scales of $\sim500$ mas ($\sim500$ pc).

\subsection{HI Absorption}
\label{HI}

\begin{figure}
\centering
\scalebox{0.8}{\includegraphics{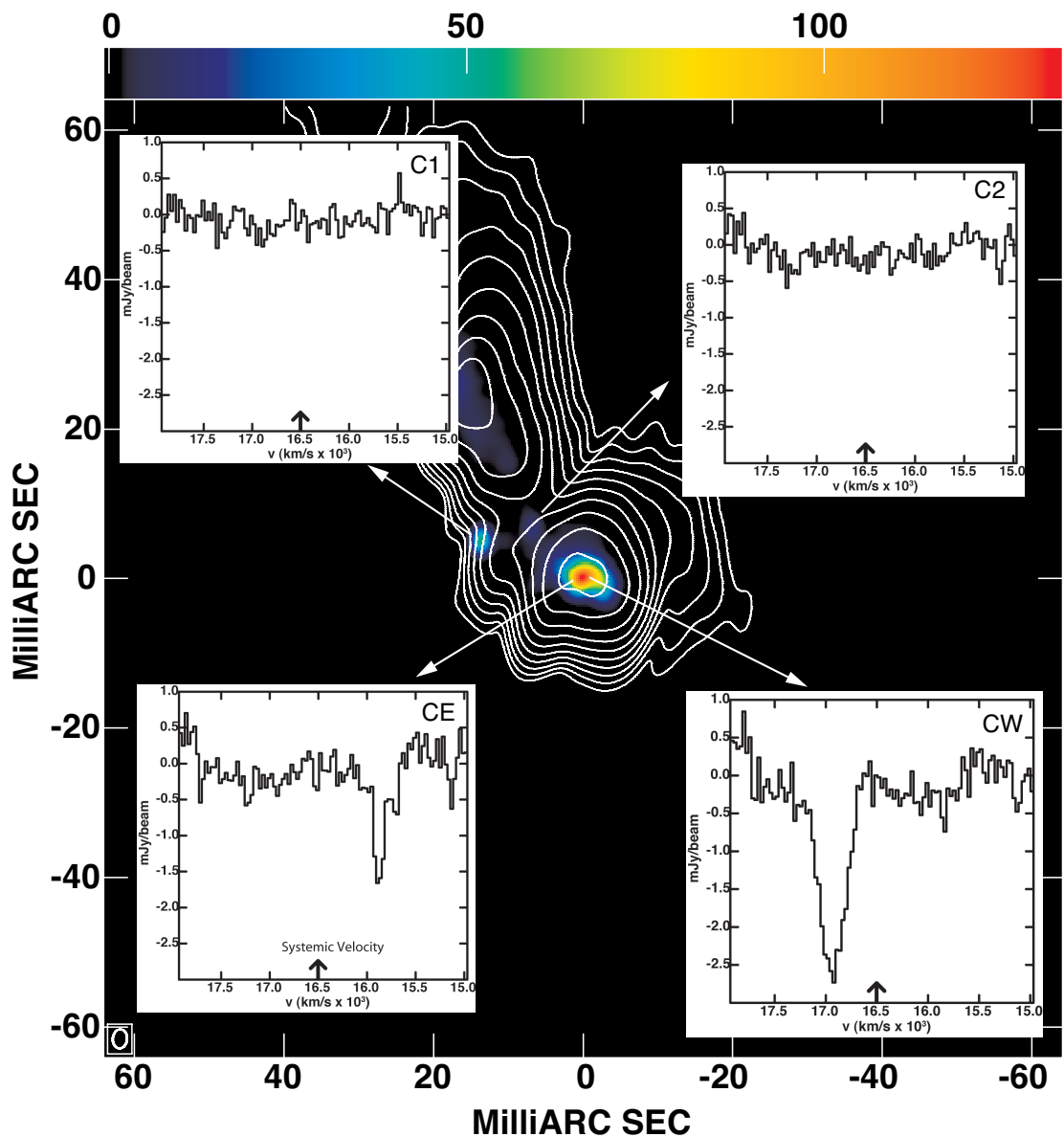}}
\vspace{-3cm}
\caption{HI absorption profiles taken from four regions of the source (components C1, C2, CE, and CW). The arrow in the center of each spectrum shows where the systemic velocity is \citep{Rod06}, and the velocity resolution is 15 km s$^{-1}$. The rms noise in a single channel is 0.14 mJy/beam. The contours are taken from the 2007 Global VLBI observations at 1.3 GHz and are set at 10$\sigma$, increasing by a factor of 2 thereafter; and the color scale image is taken from the 2005 VLBA observations at 5 GHz.}
\label{Spectra}
\end{figure}

Figure \ref{Spectra} shows HI absorption profiles taken from four regions of the source. The continuum has been subtracted from the spectra using the task UVLSF in AIPS, which removes a continuum model from the \textit{u},\textit{v} data of all channels. The contours are taken from the 2007 Global VLBI observations at 1.3 GHz and the color scale from the 2005 VLBA observations at 5 GHz. 
Two absorption lines are evident, which appear to be at two different locations toward the southern jet. From this point on we will designate CW the western component, where we see the stronger line; and CE the eastern component, where we see the weaker line. Measurements of these lines are given in Table \ref{Gaussians}, where we also show the peak opacity and the column density for both components, which we calculated according to,
\begin{eqnarray}
\label{NH}
N_{HI}(\text{cm}^{-2})&=&1.8224\times10^{18}T_S(\text{K})\int_{-\infty}^{\infty} \tau(v)\,dv(\text{km s$^{-1}$} ) \nonumber\\
 &\sim& 1.8224\times10^{18}T_S(\text{K}) \Delta v(\text{km s$^{-1}$}) \sum \tau(v), \nonumber\\
\end{eqnarray}
\noindent
where $T_S$ is the spin temperature, $\Delta v$ is the velocity resolution, and the summation is over the channels where we detect absorption for CW and CE respectively.

\begin{figure}
\centering
\scalebox{1.3}{\includegraphics{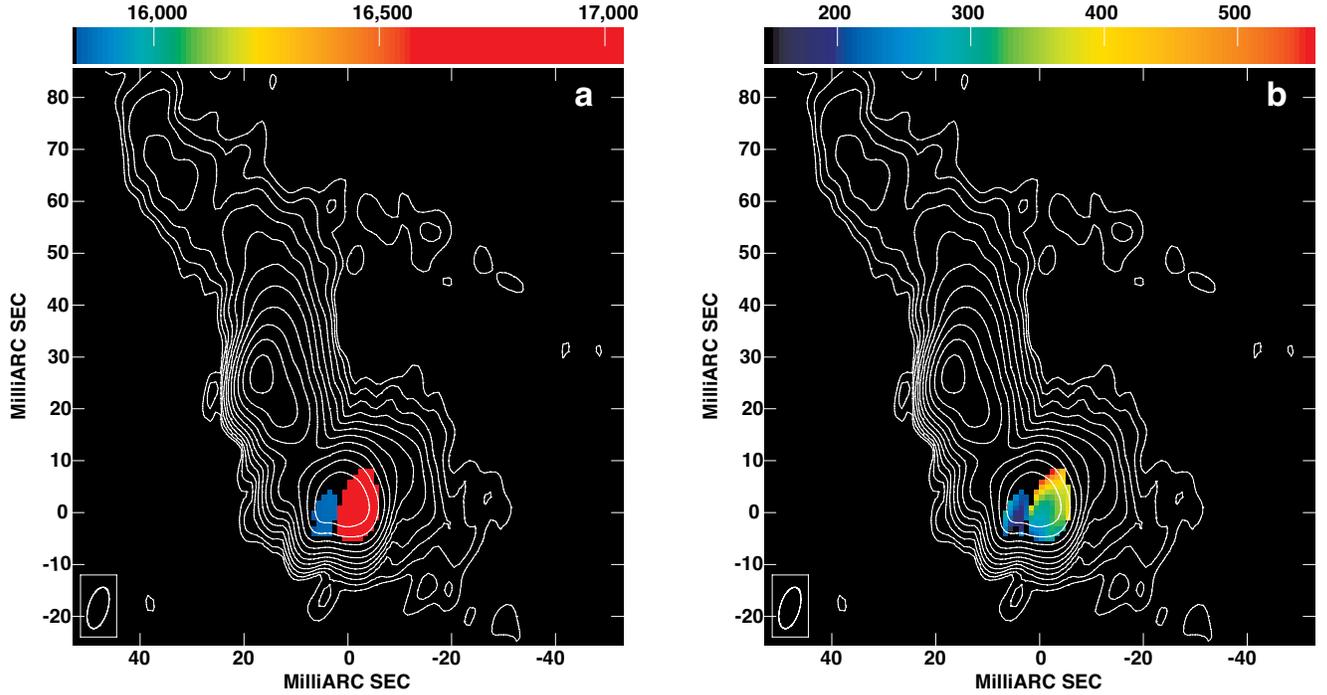}}
\vspace{2cm}
\caption{(a) Map of the central velocity and (b) width of the HI absorption profiles in the source. Both maps were generated fitting Gaussian functions at each pixel where at least a 3$\sigma$ detection of a line was found. The color scale in both images is in units of km s$^{-1}$. The contours are taken from the 2007 Global VLBI observations at 1.3 GHz and are set at 3$\sigma$, increasing by a factor of 2 thereafter.}
\label{central+width}
\end{figure}

We determined the central velocity of the HI absorption lines to be 16,927 $\pm$ 7 km s$^{-1}$ and 15,856 $\pm$ 9 km s$^{-1}$ for CW and CE respectively. From the most recent reported redshift for 0402+379, as measured from optical emission lines \citep{Rod06}, we find that the observed line for CW is redshifted by 370 $\pm$ 10 km s$^{-1}$ and for CE is blueshifted by 700 $\pm$ 10 km s$^{-1}$ from the systemic velocity of the source (16,558 $\pm$ 3 km s$^{-1}$). No absorption was found at either C1 or C2 (see Figure \ref{Spectra}). We calculate a limit on the HI opacity at the location of C1 and C2, and obtain $<$ 0.17 and $<$ 0.03 repectively. Figure \ref{central+width} shows a map of both the central velocity and width of the HI absorption profiles over the source.

\begin{figure}
\centering
\scalebox{0.6}{\includegraphics{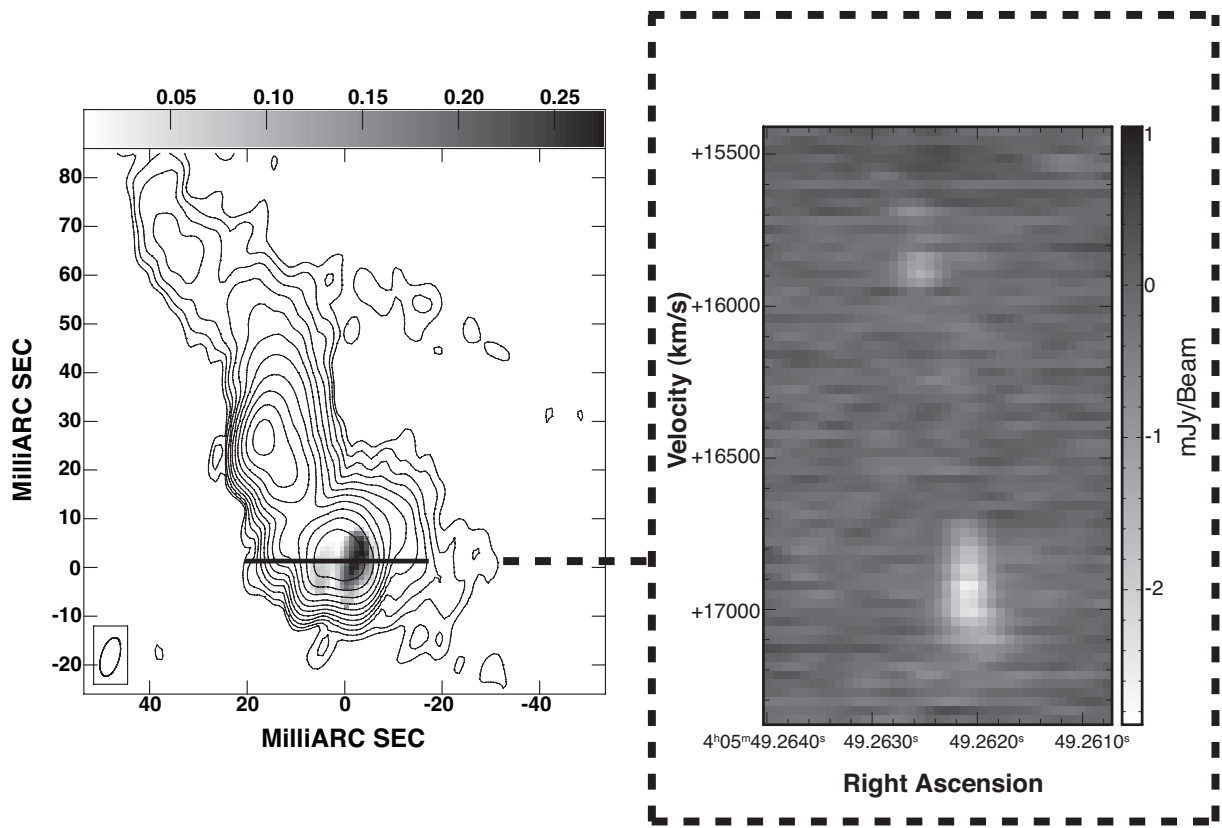}}
\vspace{2cm}
\caption{Velocity slice of the continuum-subtracted cube described in \S\ref{HI} (right panel), accompanied by the integrated HI opacity distribution over the source (left panel), in which we indicate where the slice was taken. The map was generated by combining a continuum image of the source with the continuum-subtracted cube. To generate the opacity distribution shown we did a one-dimensional zeroth moment fitting to each row, requiring a continuum emission of at least 0.12 mJy and a line emission of at least 0.75 mJy (corresponding to $\sim5\sigma$ respectively). The contours are taken from the 2007 Global VLBI observations at 1.3 GHz and are set at 3$\sigma$, increasing by a factor of 2 thereafter.}
\label{Opacity}
\end{figure}

\begin{figure}
\centering
\scalebox{0.4}{\includegraphics{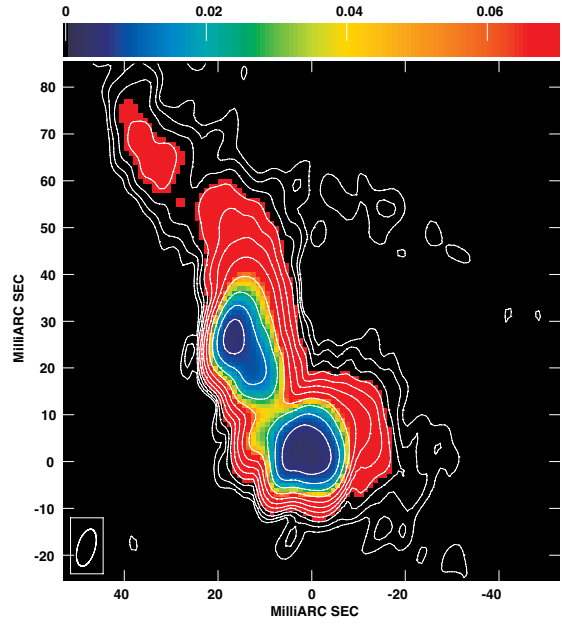}}
\vspace{2cm}
\caption{Calculated lower limit on the peak opacity across the source assuming an intensity for the line of $3\sigma$. The contours are taken from the 2007 Global VLBI observations at 1.3 GHz and are set at 3$\sigma$, increasing by a factor of 2 thereafter.}
\label{opacitypeak}
\end{figure}

Figure \ref{Opacity} shows a velocity slice of the continuum-subtracted cube (right panel), accompanied by the HI opacity distribution over the source (left panel). This result shows that either the two locations where we find absorption are localized regions, rather than being part of a more extended and perhaps homogeneous structure, or our sensitivity prevented us from detecting HI absorption over a broader region. In order to explore this question we calculated the lower limit on the peak opacity across the source assuming an intensity for the line of $3\sigma$, where $\sigma$ is the rms noise in a single channel. The result is shown in Figure \ref{opacitypeak}. We see that outside the region where we detect absorption (compare with Figure \ref{Opacity}) we would need a peak opacity of at least $\sim 0.02$, comparable to the peak  opacity found for components CE and CW, in order to have a $3\sigma$ detection. The fact that we did not detect this means that the peak opacity in this region is even smaller than the lower limit found, suggesting that what we are observing are two different, localized clumps, rather than a more extended and homogeneous structure. The fact that we do not see absorption against C2 supports this scenario.

We can estimate the HI mass of the clumps from the calculated column densities (Table \ref{Gaussians}) and assuming a spherical shape, according to the following relation,

\begin{equation}
\label{mass}
m=\dfrac{4\pi}{3}r^2N_{HI}m_{H}
\end{equation}
\noindent
where $m_{H}$ is the hydrogen atom mass and $r$ is the radius of the clump estimated from the opacity distribution map shown in Figure \ref{Opacity}. For CW we used $r\sim 12$ pc and for CE $r\sim 8$ pc. We find a mass ranging from $\sim(7$ $-$ $400) \times 10^3$ $M_{\odot}$ for component CW, and $\sim(1$ $-$ $60)\times 10^3 M_{\odot}$ for component CE, assuming spin temperatures ranging from 100 K to 6000 K. These masses are bigger than typical values of HI cloud masses of $60$ $M_{\odot}$ found in the Milky Way \citep{Stil06}.

\section{Discussion}
\label{Discussion}

The results found from our 2007 Global VLBI observations at 1.3 GHz show two HI absorption lines, one blueshifted and the other redshifted with respect to the systemic velocity. The redshifted line (component CW) shows a FWHM of 300 $\pm$ 20 km s$^{-1}$, and the blueshifted line (component CE) a FWHM of 170 $\pm$ 20 km s$^{-1}$. This absorption must be from neutral hydrogen along the
line-of-sight to the central continuum source, but this gas could be close to the nucleus or far away; could be falling
in, outflowing, or in rotation. 

Assuming a virialized cloud of $\sim 10^4$ $M_{\odot}$ and a radius of 5 pc we get, from the virial theorem, a velocity for the H atoms of $\sim 3$ km s$^{-1}$. On the other hand, if we use the high end temperature of $10^4$ K the thermal speed of an H atom is $\sim 15$ km s$^{-1}$. Thus, under these assumptions, at most $\sim 20$ km s$^{-1}$ of the line widths observed for both CE and CW could be accounted for, leaving a considerable balance which could be attributed to the proximity of these components to the central parsecs of the source,  where the gravitational potential well is deepest.

Another possibility is that the broad line widths and large velocities for one or both components are due to a jet-cloud interaction. \cite{Oosterloo} studied the Seyfert 2 galaxy IC 5063, where a strong interaction between the radio jet and a molecular cloud of the interstellar medium (ISM) is ocurring at the position of the western radio lobe of the source. The most prominent absorption feature found is blueshifted over 600 km s$^{-1}$ with respect to the systemic velocity, a value that lies outside the range allowed by rotational kinematics of the large-scale HI disk. As an effect of the outflow produced by the interaction as well as the geometry of the source, there will be components moving both away and towards the observer. Since only absorption in those components in front of the radio continuum can be observed, the blueshifted component was the only one detected. Returning to our case, it is very unlikely that the motion found in both components, CW (redshifted component) and CE (blueshifted component), which are only a few parsecs apart, is due to jet-cloud interactions, since this would require a dramatic change in the way the southern jet is moving, on scales of only a few parsecs.

We consider that both CW and CE are part of the same rotating disk structure, of which we only see a couple of clumps, and for the remainder of the discussion, we  assume that this is the case and propose a model that reproduces the observations. 
We also assume that the mass of component C1 is significantly smaller than that of component C2, thus not affecting the stability of the disk. This type of binary, where the two black holes have different masses, are thought to occur more frequently, since minor galactic mergers are more common than major mergers in hierarchical models of galaxy formation \citep{Armitage02}. We realize that in a thin disk the binary would likely clear out all material within twice the radius of the binary \citep{MacFadyen08}, but perhaps this process is less efficient for a thick disk, or material is resupplied.

\subsection{Rotating Disk}
\label{RotatingDisk}

If we assume that components CW and CE are rotating about C2 in circular Keplerian orbits lying on the same plane, the only way of reproducing the measured velocities (according to our proposed model, explained below) is by requiring a very high central mass, of at least $\sim 10^{10} M_{\odot}$, in disagreement with previous estimates \citep{Rod06} as well as typical supermassive black hole masses. 

Thus, we propose a model consisting of a geometrically thick disk, rather than a geometrically thin disk, rotating in a Keplerian circular orbit, which can be part of a larger scale torus, that can reproduce the velocities that we observe from the HI absorption, at the measured locations. We assume that CW and CE are rotating in parallel planes, both offset from the plane in which C2 lies. The distances between these planes are free parameters in our model. We call $d1$ the separation between C2 and the center of the circular orbit 1, that CE is following; and $d2$ the separation between C2 and the center of the circular orbit 2, that CW is following. Figure \ref{Cartoon} is a cartoon showing the configuration of the model we are proposing. 

We assume that the rotation axis crosses component C2, which is at the systemic velocity of 16,558 $\pm$ 3 km s$^{-1}$ \citep{Rod06}, a natural assumption considering that the jets of this source seem to be emerging from C2, not from C1. Also, we suggest that the two absorption lines that are observed could be explained by the presence of two mass clumps, both part of a clumpy geometrically thick disk structure (an idea that is supported by the opacity distribution map obtained, Figure \ref{Opacity}).

We let $d1$ and $d2$ vary over a range of values and calculate, for each different combination ($d1$,$d2$) what the orientation in space of the thick disk should be in order to best reproduce the observed velocities, via a minimization routine. After obtaining the inclination of the thick disk for each pair ($d1$,$d2$), we again apply a minimization routine in order to find the combination ($d1$,$d2$) that best reproduces the measured velocities.

This same procedure was applied with different values of the central mass, which allowed us to set a lower limit on this quantity of  $\sim1.4 \times 10^8$ $M_{\odot}$ (by requiring that the difference between the velocities predicted by the model and the observed velocities was smaller than 1$\sigma$). See Appendix \ref{DiskAppendix} for more details on the model. There is no upper limit on the mass that we can find using this model, since for greater masses we can always find an inclination that reproduces very accurately the measured values (the inclination angle, with respect to the line of sight, decreases as we increase the central mass). 

\cite{Rod06} made an estimate of the central mass of 0402+379, based on an optical spectrum of the core of this radio galaxy taken with the Hobby-Eberly Telescope, which shows a red shoulder suggesting two components with a velocity separation of $\sim$300 km s$^{-1}$. At the observed projected separation between the nuclei of 7.3 pc, an orbital velocity of 300 km s$^{-1}$ implies a system mass of $1.5 \times 10^8$ $M_{\odot}$. For this particular value of the central mass, our model gives an inclination of the rotation axis of $\sim85^{\circ}$ with respect to the line of sight, thus requiring the source to be very close to the plane of the sky. A value for $d2$ of $\sim12$ pc is obtained, which sets a lower limit on the scale height of the thick disk. This value is consistent with observed parameters of supposed tori (see for example Peck et al. 1999, where a 20 pc scale height torus is estimated for the radio source 1946+708). 

Considering the size of our synthesized beam and the extension of the two regions where absorption is observed, we were not able to measure a significant velocity gradient within CW or CE. The small variations in velocity found are not reproduced by our model. However, other factors could influence these mass concentrations that we are not taken into account in our model, such as proper motions associated with each clump.

\subsection{More Constraints on the Inclination of the Disk}
\label{JetComp}

Assuming that C2 is the origin of the radio emission on parsec scales,
we can constrain the orientation of 0402+379.  In the simple beaming
model for simultaneously ejected jet components moving in opposite
directions, the arm length ratio $ D $ depends on the intrinsic speed $\beta=v/{\rm c}$ and the angle of the
twin jets to the line of sight $\theta$ \citep{Taylor&Vermeulen97}. The arm length ratio, $D$, is given by

\begin{equation}
\label{arm_ratio}
D=\frac{d_{\rm N}}{d_{\rm S}}=\left( \frac{1+\beta \cos \theta}{1-\beta \cos \theta}\right),
\end{equation}

\noindent
where the apparent projected distances from C2 (assumed to be the origin of radio emission)
are $ d_{\rm N} $ for the northern jet (approaching side) and $ d_{\rm S} $ for the southern jet (receding side). From this we get $ \beta\cos \theta = 0.4 $ \citep{Rod06}, a result that implies that the intrinsic velocity
must be at least 0.4c and $ \theta $ must be less than $66^{\circ}$. Based on the observed morphology of the source, we can assume a minimum value for $\theta$ of $\sim30^{\circ}$. According to our proposed model (see \S\ref{RotatingDisk}) these constraints on the inclination set a lower limit on the central mass of the system of $\sim7\times 10^8$ $M_{\odot}$, for $\theta=66^{\circ}$. Angles as small as $\sim30^{\circ}$ are allowed, however, they require a very large mass ($\sim10^{10}$ $M_{\odot}$).

\begin{figure}
\centering
\scalebox{0.8}{\includegraphics{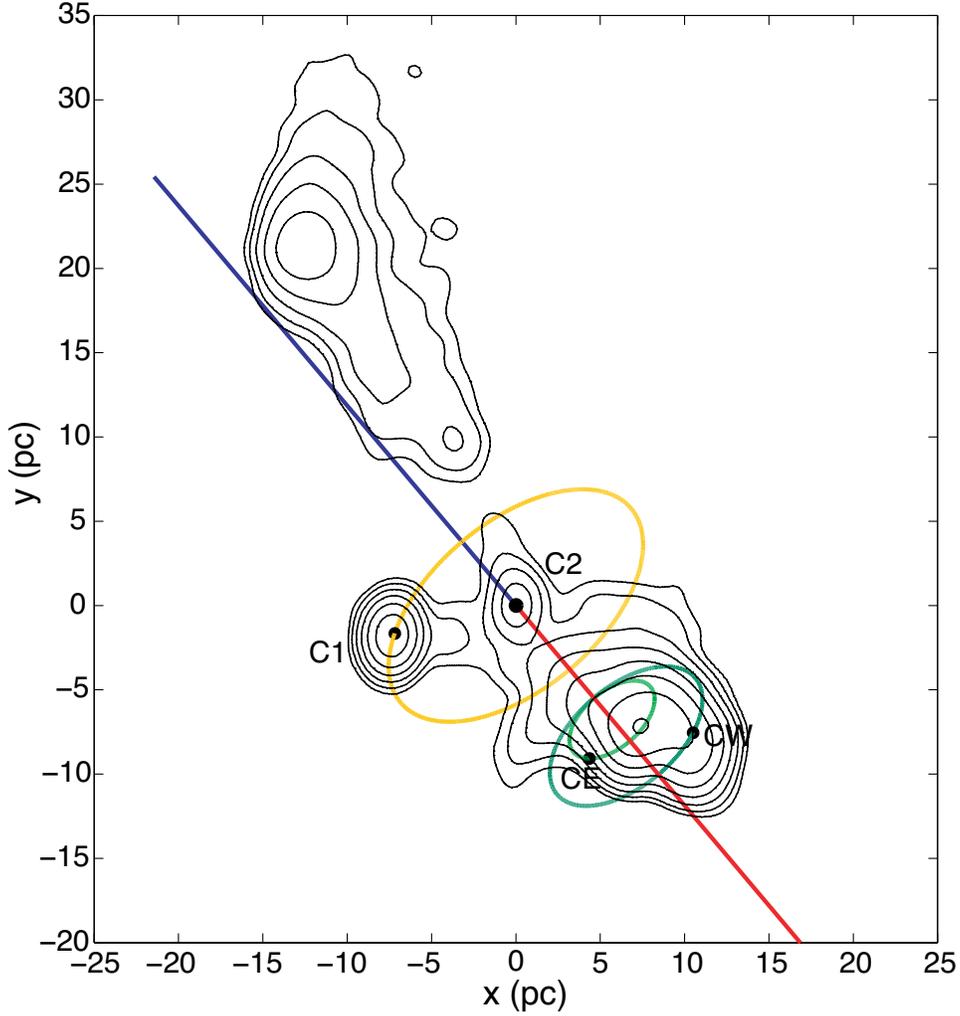}}
\vspace{2cm}
\caption{Results obtained from the rotating thick disk model, projected on the $x$-$y$ plane (RA and DEC plane), overlaid with 5 GHz contours from the 2005 VLBA observations. The blue section of the rotation axis is pointing towards us, whereas the red section is pointing away. For a mass of $7 \times 10^8$ $M_{\odot}$ an inclination angle of $\sim66^{\circ}$ between the rotation axis and the line of sight is obtained. The position of component C2, chosen as the origin, is shown, as well as the position of C1, CE and CW. The two green circular orbits drawn are the trajectories CE and CW would follow, respectively, on the rotating thick disk. The yellow circular orbit is the trajectory C1 would follow assuming C2 as the center of the orbit.}
\label{Cartoon7e8Msun}
\end{figure}

Figure \ref{Cartoon7e8Msun} is a cartoon showing the results obtained by our model for a central mass of $7 \times 10^8$ $M_{\odot}$ ($\theta=66^{\circ}$), overlaid with 5 GHz contours of 0402+379 from the 2005 VLBA observations. We can see that the rotation axis aligns very well, within $\sim10^{\circ}$, with the jet axis, which was not a constraint imposed by the model. If we asume that the center of the orbit that C1 follows is located at the position of C2, and also that component C2 is significantly more massive than C1, we obtain an orbital velocity for C1 of $\sim580$ km s$^{-1}$ (for a mass of $7 \times 10^8$ $M_{\odot}$ and $\theta=66^{\circ}$). This is consistent with the current limit from VLBI monitoring of $<26,400$ km s$^{-1}$.

\section{Conclusion}

Global VLBI observations at 1.3 GHz were performed on the radio galaxy 0402+379, the most compact supermassive binary black hole pair yet imaged. Two absorption lines were found toward the southern jet of the source, one redshifted by 370 $\pm$ 10 km s$^{-1}$ and the other blueshifted by 700 $\pm$ 10 km s$^{-1}$ with respect to the systemic velocity of the source. 

A model consisting of a thick disk, of which we only see a couple of clumps, was devoloped in order to reproduce the velocities measured from the HI absorption profiles. These clumps, components CW and CE, rotate in circular Keplerian orbits around an axis that crosses one of the supermassive black holes of the binary system in 0402+379, component C2. We found an upper limit for the inclination angle of the twin jets to the line of sight of $\theta=66^{\circ}$, which, according to the proposed model, implies a lower limit on the central mass of $\sim7 \times 10^8$ $M_{\odot}$ and a lower limit for the scale height of the thick disk of $\sim12$ pc .

{\it Facilities:} \facility{Very Long Baseline Array}, \facility{Effelsberg}, \facility{Jodrell Bank}, \facility{Westerbork}, \facility{Green Bank Telescope}, \facility{Onsala 25m telescope}

\acknowledgments 
We thank an anonymous referee for constructive comments.

The National Radio Astronomy Observatory is a facility of the National
Science Foundation operated under a cooperative agreement by
Associated Universities, Inc.

\clearpage

\begin{appendix}
\section{Rotating Disk Model}
\label{DiskAppendix}

\begin{figure}
\centering
\scalebox{1}{\includegraphics{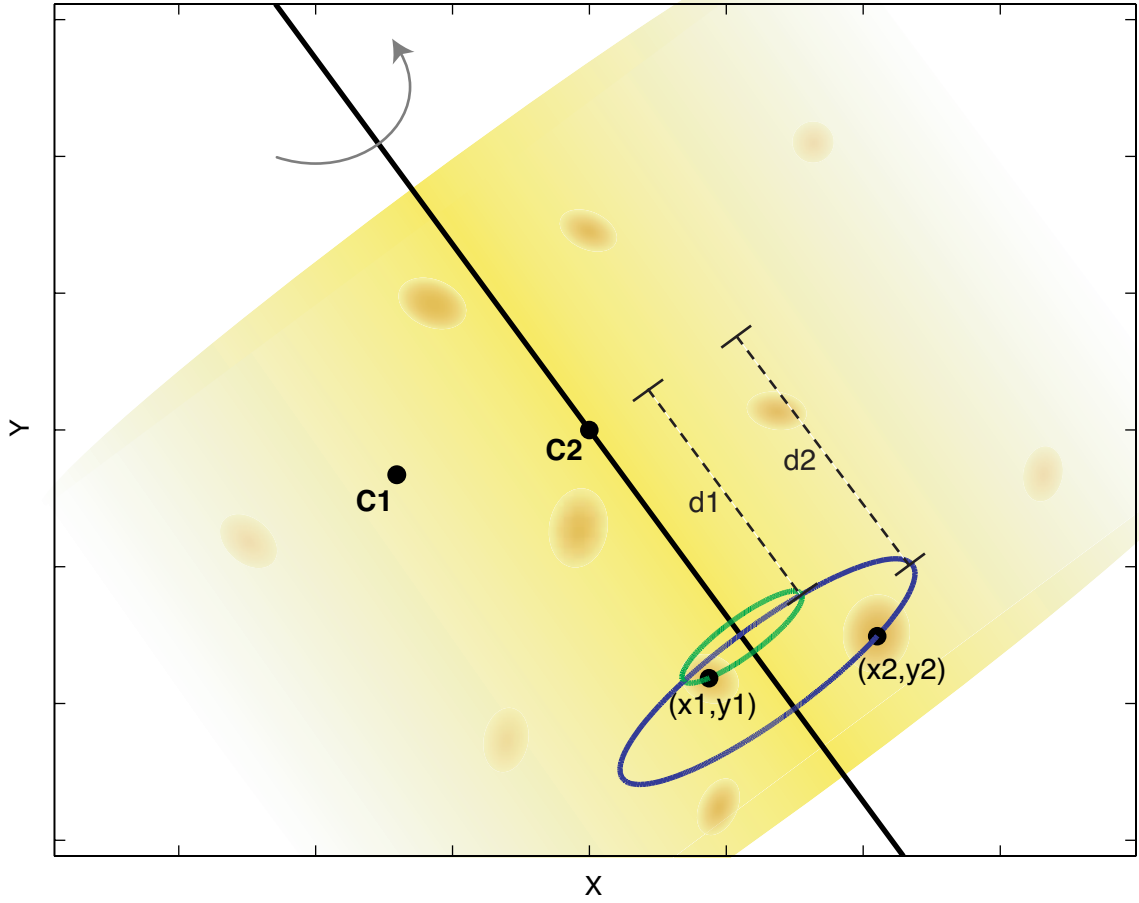}}
\vspace{2cm}
\caption{Configuration of the model we are proposing. Shown is component C2, chosen as the origin, as well as components CE, with coordinates (x1,y1), and CW, with coordinates (x2,y2). The two circular orbits drawn are the trajectories CE and CW would follow, respectively, on the rotating thick disk, shown in yellow. An inclination angle of $\sim75^{\circ}$ between the rotation axis and the line of sight was used.}
\label{Cartoon}
\end{figure}

Figure \ref{Cartoon} is a cartoon showing the configuration of the model we are proposing. We show component C2, which is chosen as the origin, as well as components CE, with coordinates (x1,y1), and CW, with coordinates (x2,y2). The two circular orbits drawn are the trajectories CE and CW would follow, respectively, on the rotating thick disk. Also, we show the projection of the system on the plane of the sky, since that is what we actually observe. 

Let us assume our disk has an arbitrary orientation in space, specified by the normal unitary vector $\widehat{n}$ = ($n_{x}$,$n_{y}$,$n_{z}$). Since $\widehat{n}$ is unitary we have,
\begin{equation}
\label{}
n_{x}^{2}+n_{y}^{2}+n_{z}^{2}=1 
\end{equation}

According to Kepler's Laws, a body moving around a mass $M$, at a radius $R$, has a velocity given by,
\begin{equation}
\label{}
V=\sqrt{\dfrac{GM}{R}}
\end{equation}

Thus, the angular velocity is given by,
\begin{equation}
\label{}
\omega=\sqrt{GM} (R)^{-3/2}
\end{equation}

If $\overrightarrow{\omega}$ is the angular velocity vector of our rotating thick disk, and $\overrightarrow{R}$ is any point on the disk, then the tangential velocity $\overrightarrow{V}$ is given by,
\begin{equation}
\label{}
\overrightarrow{V}=\overrightarrow{\omega}\times\overrightarrow{R}
\end{equation}

\noindent

where $\overrightarrow{\omega} = \omega \widehat{n}$ and $\overrightarrow{R} = \overrightarrow{r}-\overrightarrow{d}$, where $\overrightarrow{d}$ is the separation between the origin and the center of the circular orbit we are considering, and $\overrightarrow{r}$ is the position vector of the point we are considering (in our case, the coordinates of either CE or CW). Thus, we have,
\begin{equation}
\label{}
\overrightarrow{R}=(R_{x}, R_{y}, R_{z})=(x+dn_{x}, y+dn_{y}, z+dn_{z})
\end{equation}

Let the plane $x$-$y$ be the plane of the sky and $z$ be the line of sight direction. We want to know the $z$ component of the tangential velocity $\overrightarrow{V}$, because that component is the one we can measure (from our absorption lines), which is given by:
\begin{equation}
\label{}
V_{z}=\overrightarrow{V}.\widehat{z}=\omega(n_{x}R_{y} - n_{y}R_{x} )
\end{equation}

Finally, we can also use the fact that $\overrightarrow{\omega}$ and $\overrightarrow{R}$  have to be orthogonal, that is,
\begin{equation}
\label{}
\overrightarrow{\omega}.\overrightarrow{R}=0 \Longrightarrow n_{x}x + n_{y}y + n_{z}z + d=0
\end{equation}

We can now put everything together and get an expression for $V_{z}$,
\begin{equation}
\label{velocityZ}
V_{z}=\sqrt{GM} \left[ n_{x}(y + dn_{y}) - n_{y}(x+ dn_{x}) \right]  \left[ (x+ dn_{x})^{2} + (y + dn_{y})^{2} + (z + dn_{z})^{2} \right] ^{-3/4} 
\end{equation}
\noindent

where 
\begin{equation}
\label{}
z= \dfrac{-d-xn_{x}-yn_{y}}{n_{z}} \text{ \ \    and   \ \  } n_{z} = \sqrt{1 -n_{x}^{2}-n_{y}^{2}}
\end{equation}
\noindent

Thus, we have an equation that gives us the $z$ component of the tangential velocity of any point with a circular orbit in our rotating disk, in terms of the $x$ and $y$ components of the normal unitary vector of the disk, $n_{x}$ and $n_{y}$, and in terms of the $x$ and $y$ components of any point in the disk.

There are two locations in our source, where we can measure the velocity. Let us define $V_{z1,o}$ and $V_{z2,o}$ as the observed velocities of CE and CW respectively; and $V_{z1,c}$ and $V_{z2,c}$ the calculated velocities according to Equation \ref{velocityZ}. Finally let us define the function that we will minimize in order to obtain the inclination of the thick disk that reproduces the measured velocities best,
\begin{equation}
\label{}
J= \dfrac{ |V_{z1,o} - V_{z1,c}|}{|V_{z1,o}|} + \dfrac{|V_{z2,o} - V_{z2,c}|}{|V_{z2,o}|}
\end{equation}

For a particular value of the central mass of the system, $M$, we went through the following routine in order to find the inclination of the thick disk, more precisely $n_{x}$ and $n_{y}$ (since $n_{z}$ can be determined in terms of these two quantities): We calculated the function $J$ for every combination of ($n_{x}$, $n_{y}$), with the constraint that $\widehat{n}$ is unitary. By minimizing $J$ we found the best pair ($n_{x}$, $n_{y}$). Since $d1$ and $d2$ are also free parameters, we let them vary over a range, so for every combination ($d1$, $d2$) we followed the procedure mentioned above and found the best pair ($n_{x}$, $n_{y}$). After having done this for every combination ($d1$, $d2$), again by a minimization routine, we found the best pair ($d1$, $d2$). Thus, for a particular value of the central mass we were able to find the inclination of the disk that reproduced the observed velocities best. This same routine was also followed for different values of the central mass, which let us find a lower limit on this quantity of  $\sim1.4 \times 10^8$ $M_{\odot}$, by requiring that the difference between the velocities predicted by the model and the observed velocities was smaller than 1$\sigma$, with:

\begin{equation}
\label{}
\sigma = \sqrt{ \dfrac{(V_{z1,o} - V_{z1,c})^{2}}{\Delta_{1}^{2}} + \dfrac{(V_{z2,o} - V_{z2,c})^{2}}{\Delta_{2}^{2}}} 
\end{equation}
\noindent
where $\Delta_{1}$ and $\Delta_{2}$ are the errors in the measurements of $V_{z1,o}$ and $V_{z2,o}$ respectively.

\end{appendix}

\clearpage

\begin{deluxetable}{lccccc}
\tabletypesize{\scriptsize}
\tablecolumns{6}
\tablewidth{0pt}
\tablecaption{Gaussian Functions Fitted to HI Absorption Profiles in the Southern Jet.\label{Gaussians}}
\tablehead{\colhead{Component}\tablenotemark{*}&\colhead{Amplitude (mJy)}&\colhead{Central Velocity (km s$^{-1}$)}&\colhead{FWHM (km s$^{-1}$)}&\colhead{$\tau_{peak}$}&\colhead{$N_{HI}$ (cm$^{-2}$) }}
\startdata
CW  & 2.8 $\pm$ 0.1 & 16,927 $\pm$ 7  & 300 $\pm$ 20 & 0.025 $\pm$ 0.001  & (1.303 $\pm$ 0.006) $\times 10^{21}$  \tablenotemark{a} \\ 
  &  &  &  &  & (7.82 $\pm$ 0.03) $\times 10^{22}$   \tablenotemark{b}\\ 
CE  & 1.5 $\pm$ 0.2 & 15,856 $\pm$ 9  & 170 $\pm$ 20 & 0.018 $\pm$ 0.002 &  (4.1 $\pm$ 0.1) $\times 10^{20}$  \tablenotemark{a}\\
  &  &  &  &  & (2.44 $\pm$ 0.07) $\times 10^{22}$   \tablenotemark{b}\\ 
\enddata
\tablenotetext{*}{CW and CE refer to the western and eastern jet components where we find absorption lines respectively.}
\tablenotetext{a}{Assuming a spin temperature of 100 K}
\tablenotetext{b}{Assuming a spin temperature of 6000 K}

\end{deluxetable}

\clearpage

\end{document}